\documentclass{article}
\usepackage{spconf,amsmath}
\usepackage{amsthm}
\usepackage{amsfonts}
\usepackage{amssymb}
\usepackage{mathrsfs}
\usepackage{mathtools}
\usepackage[english]{babel}
\usepackage{float}
\usepackage[toc,page]{appendix}
\usepackage[pdftex]{graphicx}
\usepackage{epstopdf}
\usepackage{amsmath}
\usepackage{color}
\usepackage{multirow}
\usepackage{graphicx}
\usepackage{IEEEtrantools}
\usepackage{bm}
\usepackage{balance}
\usepackage{microtype}
\usepackage[caption=false,font=footnotesize]{subfig}
\usepackage{cite}
\usepackage{comment}

\numberwithin{equation}{section}

\usepackage[switch]{lineno}
\usepackage[named]{algo}
\usepackage[noend]{algpseudocode}
\usepackage{algorithm}
\allowdisplaybreaks 

\ninept

\newcommand{\dddag}{%
  \mathbin{\vbox{\offinterlineskip\ialign{%
    \hfil##\hfil\cr
    \small{$\dagger$}\cr
    \noalign{\kern-0.6ex}
    \small{$\ddagger$}\cr
}}}}

\title{Evaluation of Orthogonal Chirp Division Multiplexing for Automotive Integrated Sensing and Communications} 
	
\name{Sangeeta Bhattacharjee$^{\dag}$, Kumar Vijay Mishra$^{\ddag}$, Ramesh Annavajjala$^{\dddag}$ and Chandra R. Murthy$^{\dag}$}
\address{$^{\dag}$Indian Institute of Science, Bangalore 560012 India\\
$^{\ddag}$United States CCDC Army Research Laboratory,
Adelphi, MD 20783 USA\vspace{-4pt}\\ 
$^{\dddag}$Northeastern University, Boston, MA 02115 USA}

\begin{document}
\setlength{\abovedisplayskip}{3pt}
\setlength{\belowdisplayskip}{3pt}

\maketitle

\begin{abstract}
We consider a bistatic vehicular integrated sensing and communications (ISAC) system that employs the recently proposed orthogonal chirp division multiplexing (OCDM) multicarrier waveform. As a stand-alone communications waveform, OCDM has been shown to be robust against the interference in time-frequency selective channels. In a bistatic ISAC, we exploit this property to develop efficient receive processing algorithms that achieve high target resolution as well as high communications rate. We derive statistical bounds for our proposed \textit{S}eq\textit{u}e\textit{n}tial symbol \textit{d}ecoding and radar p\textit{a}rameter \textit{e}stimation (SUNDAE) algorithm and compare its competitive performance with other multicarrier waveforms through numerical experiments. 
\end{abstract}

\vspace{-8pt}
\section{Introduction}
\label{sec:intro}
The advent of cellular communications technologies and novel radar applications has led to the electromagnetic spectrum becoming a contested, scarce resource \cite{mishra2019toward,petropulu2020dfrc,liu2020joint}. As a result, there have been a slew of developments in system engineering and signal processing techniques to enable optimal spectrum sharing between radar and communications \cite{paul2016survey}. Broadly, spectrum-sharing technologies are classified into three major categories: coexistence \cite{wu2021resource}, cooperation \cite{bicua2018radar}, and co-design \cite{liu2020co}. While coexistence comprises management of interference from different radio emitters in the legacy radar and communications systems \cite{aubry2015new}, the spectral cooperation technique involves exchange of additional information between the two systems to enhance their respective performances \cite{chiriyath2017radar}. The co-design or integrated sensing and communications (ISAC), on the other hand, requires designing new systems and waveforms to efficiently utilize the spectrum through use of common transmit-receive hardware and/or waveforms \cite{duggal2020doppler}. 

Common ISAC applications include modern sensor-driven vehicles and drones \cite{dokhanchi2019mmwave,elbir2021terahertz,yu2021noncontact}. These systems not only integrate various remote sensing modalities to impart autonomy but also increasingly provide software-managed in-vehicle functions and inter-vehicular communications \cite{gerla2014internet,bengler2014three,kong2017millimeter}. Nowadays, millimeter-wave (mmWave) automotive systems that operate at $77$ GHz are preferred for both sensing and communications. The mmWave band offers contiguous wide bandwidths of the order of $4$-$7$ GHz that enable very high radar range resolution as well as high data rates \cite{mishra2019toward,dokhanchi2019mmwave,duggal2020doppler}. In this paper, we focus on mmWave automotive ISAC scenarios.

For efficient spectrum utilization, the ISAC systems often employ a common waveform for remote target detection and embedding communications symbols \cite{mishra2019toward}. In this context, lately, multi-carrier waveforms \cite{levanon_mf_radar_2000} have emerged as a practical alternative \cite{donnet2006combining,bica2016generalized,metcalf2015analysis}. The upshot is that the same transceiver structures is used for both radar and communications functionalities and additional degrees-of-freedom are available to deal with dense spectral usage, as in the case of moving vehicles in automotive scenario. A general drawback of multi-carrier radar waveforms is their time-varying envelope leading to an increased peak-to-average-power-ratio (PAPR), which makes it difficult to efficiently use the high power amplifiers \cite{wulich2005definition}. 

Among the existing multi-carrier waveforms, the conventional orthogonal frequency-division multiplexing (OFDM) has been extensively investigated for ISAC applications\cite{sturm2011waveform,metcalf2015analysis,dokhanchi2019mmwave}. Recently, orthogonal time frequency space (OTFS) modulation \cite{hadani2017orthogonal}, where the data symbols are modulated in the delay-Doppler domain, was investigated for ISAC \cite{gaudio2020effectiveness} because of its resilience to multipath components undergoing potentially different Doppler shifts. Recently, chirp spread spectrum waveforms \cite{ouyang2017chirp,vangelista2017frequency} have emerged as viable candidates for robust communications that also maximize spectral efficiency. In this context, orthogonal chirp division multiplexing (OCDM) \cite{ouyang2016orthogonal} waveform is shown to have better performance in multipath channels and robustness against interference than OFDM while offering the same PAPR. In OCDM, the discrete Fresnel transform (DFnT) is applied analogously as the discrete Fourier transform (DFT) in OFDM; each OCDM symbol is linearly spread over a fixed group of orthogonal chirps in the Fresnel domain before it is transmitted in the time domain. The DFnT could be computed using DFT \cite{bhandari2019shift} and, therefore, OCDM could be easily synthesized in digital circuits. In this paper, we present an OCDM transceiver design for a bi-static \cite{dokhanchi2019mmwave} ISAC over time-frequency selective channels. 

In particular, we propose a pilot-aided frequency-domain least squares (LS) channel estimation scheme by exploiting the circulant property of DFnT. We estimate the data symbols under negligible intercarrier interference (ICI) and then use them for target parameter estimation. Our proposed \textit{a}eq\textit{u}e\textit{n}tial symbol \textit{d}ecoding and p\textit{a}rameter \textit{e}stimation (SUNDAE) algorithm applies a maximum likelihood (ML) estimator on the decoded data symbols obtained from the communications channel to estimate the bi-static ranges and Doppler velocities of targets. Numerical experiments highlight that the resulting bit error rate (BER) is very close to that of perfect channel state information (CSI) for high communications signal-to-noise ratios (SNRs). Our comparison of the ISAC performance of OCDM shows the former is a promising alternative to OFDM and OTFS.

In the next section, we introduce the system model for OCDM-based bi-static ISAC. We present algorithms for decoding the communications symbols and estimating the target parameters in Section~\ref{sec:rx_proc}. In Section~\ref{sec:numexp}, we validate our approach and compare against existing waveforms through numerical experiments. We conclude in Section~\ref{sec:summ}. 

\vspace{-8pt}
\section{System Model}
\label{sec:sysmod}
Consider a bistatic automotive ISAC system, where an integrated radar-communications signal is emitted by a transmit (Tx) vehicle to a receive (Rx) vehicle for information transfer. This signal is also reflected-off the targets-of-interest and then captured by the Rx. The Tx  waveform is an OCDM signal, which multiplexes a bank of chirps in the same time-period and bandwidth. 
A Tx signal frame comprises $N$ temporal symbols and $M$ sub-chirps. The amplitude and phase of each chirp are used for modulating the communications bits that occupy a total bandwidth $B=M \Delta f$, where $\Delta f$ is the bandwidth of each chirp and $M$ is an even positive integer. With such a chirp basis, the baseband Tx OCDM signal is \cite{ouyang2016orthogonal}
\begin{align} \label{eq:ocdm symbol}
\begin{split}
& \hspace{3cm} \mathbf{S}=\mathbf{\Phi}_M^H \mathbf{X} \\
&\hspace{-0.3cm}\text{where} \
\mathbf{\Phi}_M=\frac{1}{\sqrt M} \mathbf{\Theta}_1 \mathbf{F}_M \mathbf{\Theta}_2,
\text{with} \ \ \mathbf{F}_M = \left[\frac{1}{\sqrt M}  e^{\mathrm j\frac{2\pi}{M}mn}\right],  \\ & \mathbf{\Theta_1}=\text{diag} \lbrace \Theta_{1,0}; \cdots;\Theta_{1,M-1}\rbrace, \ \  \Theta_{1,m}=e^{-\mathrm j\frac{\pi}{4}}e^{\mathrm j\frac{\pi}{M}m^2}, \\ & \mathbf{\Theta_2}=\text{diag} \lbrace \Theta_{2,0}; \cdots;\Theta_{2,N-1}\rbrace, \ \  \Theta_{2,n}=e^{\mathrm j\frac{\pi}{M}n^2}.
\end{split}
\end{align}
Here, $\mathbf \Phi_M^H \in \mathbb{C}^{M \times M}$ denotes the inverse DFnT (IDFnT) of order $M$ and $\mathbf{X} \in \mathbb{C}^{M \times N}$ is the matrix of data symbols, $x_{m,n}$, where $m=\lbrace 0, \cdots, M-1 \rbrace$ is the chirp index and $n=\lbrace 0, \cdots, N-1\rbrace$ is the symbol index. The IDFnT is the product of the DFT matrix $\mathbf{F}_M$ and additional quadratic phases. Furthermore, the matrix $\mathbf{\Phi}_M$ is circulant. Hence, using the eigen-decomposition property, \eqref{eq:ocdm symbol} becomes
\begin{align} \label{eq:ocdm eigen}
\begin{split}
\mathbf{S}&=\mathbf{F}_M^H \mathbf{\Gamma}^H\mathbf{F}_M \mathbf{X}=\mathbf{F}_M^H\mathbf{Z}, \\
\end{split}
\end{align}
where $\mathbf{\Gamma}^H=\mathbf{F}_M\mathbf{\Phi}_M^H\mathbf{F}_M^H$ and the matrix $\mathbf{G}=\mathbf{\Gamma}^H\mathbf{F}_M \in \mathbb{C}^{M \times M}$ transforms the input data symbols $\mathbf{X}$ into scaled frequency domain symbols $\mathbf{Z}=\mathbf{GX}$. The $\mathbf \Gamma \in \mathbb {C}^{M \times M}$ is a  diagonal matrix, where the $m$-th diagonal entry $\Gamma(m)$ is the $m$-th eigen value of $[\mathbf{\Phi}_M]_m$, and also corresponds to the root Zadoff-Chu sequences as  \cite{ouyang2016orthogonal}
\begin{align} \label{eq:gamma}
    \Gamma(m)=e^{-\mathrm j \frac{\pi}{M}m^2}, \ \forall m, \ M \equiv 0 \ (\textrm{mod} 2).
\end{align}

\begin{figure}[t]
    \includegraphics[scale=0.4]{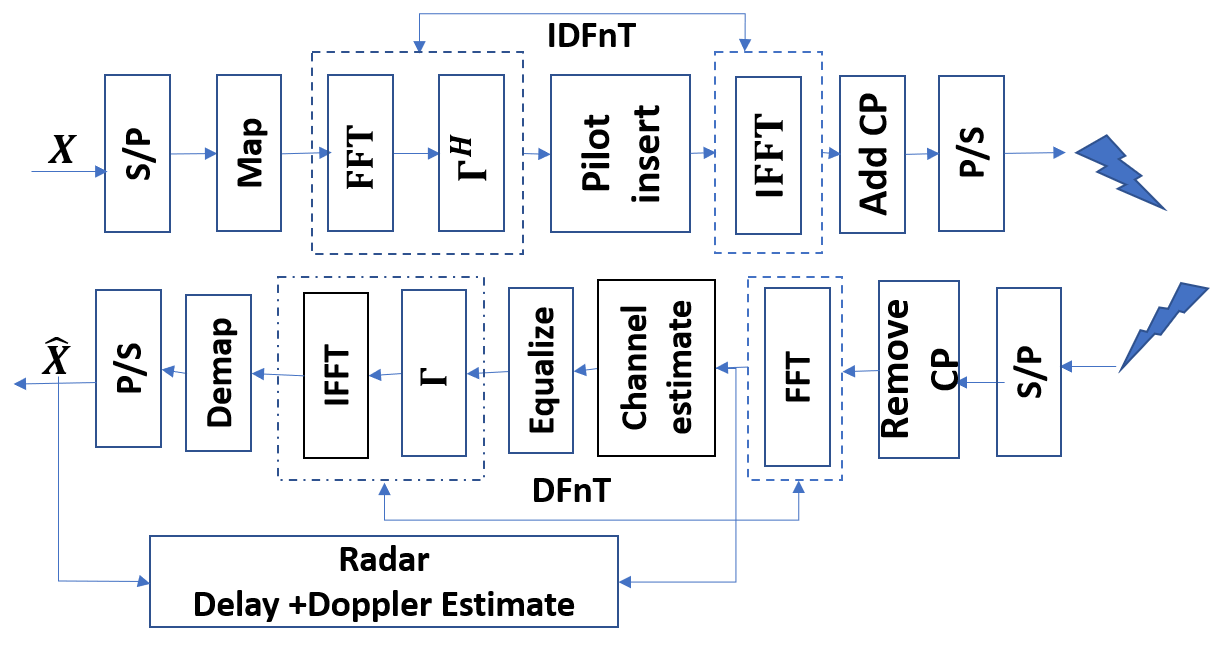}
\caption{Proposed OCDM-based ISAC transceiver that exploits the DFnT properties to integrate with the legacy OFDM systems.\vspace{-8pt}
}
\label{fig:OCDM Scheme}
\vspace{-0.5cm}
\end{figure}

The circulant property of DFnT \eqref{eq:ocdm eigen} allows the OCDM modulator to be integrated with a conventional OFDM modulator, using an additional DFT-based precoding operation $\mathbf{G}$.
In this context, Fig. \ref{fig:OCDM Scheme} presents a transceiver design for OCDM-based ISAC system that is readily integrated with the legacy OFDM systems.  

At the Tx, the data symbols are first mapped into $M$ sub-carriers and subsequently transformed to frequency domain as in \eqref{eq:ocdm eigen}. The pilot symbols are then inserted and followed by an IDFT operation that yields $N$ time-multiplexed OCDM symbols $\mathbf{S}$ in the baseband. A cyclic prefix (CP) of length $L_{\text{cp}}$ is added to each modulated OCDM symbol for mitigating inter-symbol interference (ISI) as
\begin{align}
\bar{\mathbf{S}}
=
\begin{bmatrix}  \label{eq:CP add}
\mathbf 0_{L_{\text{cp}} \times (M-L_{\text{cp}})} & \mathbf{I}_{L_{\text{cp}}} \\
\multicolumn{2}{c}{\mathbf I_{M} }\\
\end{bmatrix}\mathbf{S}. 
\end{align}
The resulting symbols are then serialized, passed through a pulse shaping filter, up-converted and transmitted. 
Define $T_0=T_{\text{cp}}+T$ as the OCDM symbol duration, where $T_{\text{cp}}$ and $T$ denote the CP and data symbol duration, respectively. The time-domain OCDM signal is
\begin{align}
\begin{split}
\label{eq:ocdm time domain}
\bar{s}(t) &=\sum\limits_{n=0}^{N-1} \sum\limits_{m=0}^{M-1} x_{m,n} e^{\mathrm{j}\frac{\pi}{4}}
e^{\left(\frac{-\mathrm{j}\pi M}{T^2}\left(t-nT_0-T_{\text{cp}}-\frac{mT}{M}\right)^2\right)}  e^{\mathrm j 2 \pi f_c t}\\ & \hspace{5cm}\textrm{rect}(t-nT_0),
\end{split}
\end{align}
where $f_c$ is the carrier frequency and  $\textrm{rect}(t)=\left\{\def\arraystretch{1.2}\begin{tabular}{@{}l@{\quad}l@{}}
  $1$ &\; $0 \le t \le T_0$ \\
  $0$ & \;\textrm{otherwise.}
\end{tabular}\right.$

The OCDM signal passes through a frequency-selective time-varying  $P$-path fading channel, whose impulse response is 
\begin{equation} \label{eq:generic channel}
h(t,\tau)=\sum \limits_{p=0}^{P-1} h_p \exp(\mathrm j 2 \pi \vartheta_p t) \delta (\tau-\tau_p),
\end{equation}
where $h_p$, $\tau_p$ and $\vartheta_p$ are the channel gain, delay, and Doppler shift of the $p^{\text{th}}$ path, respectively. This model applies to both radar and communications channels: while $h_p$ specifies the path loss coefficient for the communications channel, it characterizes the target characteristics such as the complex reflectivity or radar cross section. Further, $P$ may denote the number of multi-path links in the communications channel or non-fluctuating point-targets following the Swerling-0 target model \cite{skolnik2008radar}. 

The Tx signal \eqref{eq:ocdm time domain} is reflected back by the targets and these echoes are received by the ISAC Rx, which aims to recover the following information about the targets: time delay $\tau_{p}=\tau^{(1)}_{p}+\tau^{(2)}_{p}$, which is linearly proportional to the target's bi-static range $r_p = c\tau_p$, (superscripts (1) and (2) denote variable dependency on the Tx-target and target-Rx paths, respectively), Doppler frequency $\vartheta_p=\vartheta^{(1)}_{p}+\vartheta^{(2)}_{p}$, proportional to the target's radial velocity $v_p= c\frac{\vartheta_p}{ f_c}$ and $c$ is the speed of light. 

We use the superscripts $(\cdot)=\lbrace \text{com}, \text{\text{rad}} \rbrace$ to distinguish the parameters of the communications and radar channels, respectively. The Rx vehicle receives the OCDM signal  \eqref{eq:ocdm time domain} from the Tx over a doubly spread communications and radar channel \eqref{eq:generic channel} as
\begin{align} \label{eq:ocdm com output}
\bar{y}^{\text{com}}(t)&= \sum \limits_{c=1}^{P^{\text{com}}} h_c^{\text{com}} \bar{s}(t-\tau_c^{\text{com}})e^{\mathrm{j}2\pi \vartheta_c^{\text{com}}(t-\tau_c^{\text{com}})} +w(t), \\
\label{eq:ocdm rad output}
\bar{y}^{\text{rad}}(t)&= \sum \limits_{p=1}^{P^{\text{rad}}} h_p^{\text{rad}} \bar{s}(t-\tau_p^{\text{rad}}) e^{j2\pi \vartheta_p^{\text{rad}}(t-\tau_p^{\text{rad}})} +w(t),
\end{align}
where $w(t)$ represents the additive white Gaussian noise (AWGN). Define a generic antenna gain $g_a$. The SNR of the Rx communications and radar signals are, respectively,
\begin{align} \label{eq:SNR}
\text{SNR}^{\text{com}}=\sum \limits_{c=1}^{P^{\text{com}}} |h_c^{\text{com}}|^2 \frac{P_{\text{avg}}}{\sigma_w^2}, \ \ \text{SNR}^{\text{rad}}= \sum \limits_{p=1}^{P^{\text{rad}}} |h_p^{\text{rad}}|^2 \frac{P_{\text{avg}}}{\sigma_w^2},
\end{align}
where $P_{\text{avg}}$ and $\sigma_w^2$ denote the average Tx power and noise variance, respectively. Assume $\frac{1}{MN}\sum \limits_{n=0}^{N-1} \sum \limits_{m=0}^{M-1} \mathbb E[|x_{m,n}|]^2 \leq P_{\text{avg}}$ and  $h_c^\text{com} \sim \mathcal{CN}(0,1)$. 
 The CP duration $T_{\text{cp}}$ is chosen such that the maximum delay $\tau_{\max}=\max(\tau^{\text{\text{com}}}_{\max},\tau^{\text{\text{rad}}}_{\max})< T_{\text{cp}}<T$. Also, the subcarrier spacing is set to be larger than the maximum Doppler shift, i.e., $\vartheta_{\max}=\max(\vartheta^{\text{com}}_{\max},\vartheta^{\text{rad}}_{\max}) < \Delta f$.

\vspace{-8pt}
\section{Receiver Processing}
\label{sec:rx_proc}
The radar signal \eqref{eq:ocdm rad output}  arrives at a much higher delay than the communications signal \eqref{eq:ocdm com output}. Further, $\text{SNR}^{\text{com}} \gg \text{SNR}^{\text{rad}}$ because of higher attenuation of radar signal from target scattering. It is reasonable to assume that both signals appear as uncorrelated noise to each other at the receiver. The receiver first decodes the communications symbols and then uses the decoder output for target parameter estimation. 

The signal \eqref{eq:ocdm com output} is down-converted and sampled at $t=nT_0+T_{\text{cp}}+m\frac{T}{M}$ instants. The inter-symbol interference (ISI) is avoided with CPs. The $n$-th received symbol after removing the CP and subsequently taking an $M$ point DFT at the Rx (Fig.~\ref{fig:OCDM Scheme}) is 
\begin{align} \label{eq:OCDM DFT}
[\mathbf{Y}_f^{\text{com}}]_n=[\mathbf{H}_f]_n[\mathbf{Z}]_n + [\mathbf{W}]_n, \ \forall n,
\end{align}
where $[\mathbf{H}_f]_n=\mathbf{F}_M [\mathbf{H}]_n \in \mathbb{C}^{M \times M}$ is the channel frequency response for the $n$-th symbol, which corresponds to the discrete channel impulse response matrix $[\mathbf{H}]_n $ =$h(t,\tau)|_{t=nT+m\frac{T}{M}} \in \mathbb{C}^M$ from \eqref{eq:generic channel} for the $n$-th symbol. Further, $\mathbf{Z}=\mathbf{F}_M \mathbf{S}$, and $\mathbf{W}=\mathbf{F}_M\bar{\mathbf{W}}$, where $\bar{\mathbf{W}}=w(t)|_{t=nT+m\frac{T}{M}}$.

We now present a novel technique to estimate the frequency-domain channel $\mathbf{H}_f$ using pilot symbols and, subsequently, the data symbols ${\mathbf{X}}$ by estimating ${\mathbf{Z}}$ in \eqref{eq:OCDM DFT}.

\subsection{Pilot-based frequency-domain channel estimation} 
\begin{figure}[t]
\includegraphics[scale=0.4]{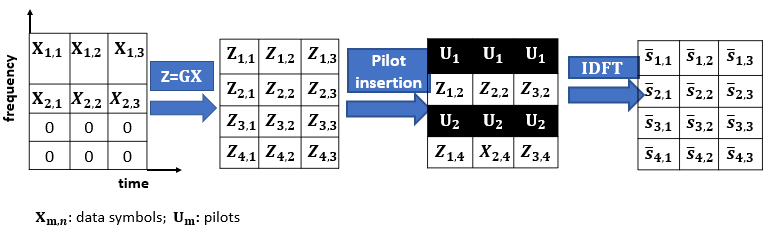}
\caption{Schematic of the OCDM transmitter with data and pilots.\vspace{-8pt}
}
\label{fig:pilot arrangement}
\end{figure}

Consider a comb-type pilot arrangement, wherein $M'_P$ pilot symbols are inserted periodically every $L$ locations apart, uniformly across all OCDM symbols (Fig.~\ref{fig:pilot arrangement}). In other words, $M$ subcarriers are grouped into $L$ adjacent subcarriers, where $L=M/M'_P$ and the first subcarrier within each group is loaded with pilot symbol $U(k)$, $k=0, \cdots, M'_P-1$. Note that the pilot insertion stage in our proposed scheme resembles that in an OFDM system (before the IDFT operation). This ensures that the DFT-precoded OCDM data symbols are directly passed through an existing OFDM modulator and demodulator block. However, in our proposed OCDM system, this is achieved by replacing the frequency domain symbols $\mathbf{Z}=\mathbf{GX}$ at pilot locations. Since each element $m$ of the DFT output vector $[\mathbf{Z}]_n$ has dependence on only the $m$-th row of $\mathbf{F}_M$, this approach works well by keeping $\mathbf{Z}$ intact at the data locations with the forward DFT transform. 

The recovery of data symbols at the receiver with an IDFT operation using straightforward insertion of pilots would result in loss of information symbols. Keeping this in mind, we load first $M-M'_P$ subcarriers with data symbols and perform zero padding of length $M'_P$ for the remaining $M'_P$ subcarriers (Fig.~\ref{fig:pilot arrangement}). The zero padding implies that each input symbol vector $[\mathbf{X}]_n$, corresponding to OCDM input symbol $n$ is transformed by the first $M-M'_P$ columns of $\mathbf{F}_M$, i.e., each DFT output symbol is given by $[\mathbf{Z}]_n=\sum \limits_{i=0}^{M-M'_P} X(i,n)e^{\mathrm j \frac{2\pi}{M}im}$. Thus, each element of $\mathbf{Z}$ at the data and pilot location is
\begin{align} \label{eq:DFT submatrix}
\begin{split}
\mathbf{Z}(kL+l,n)&=\begin{cases}
\mathbf{U}(k), \ l=0, \ k=\lbrace 0, \cdots, M'_P-1 \rbrace;\\
\mathbf{\Gamma}^H(kL+l)\mathbf{F}_M(kL+l,q) \mathbf{X}(q,n), \\ l=\lbrace 1, \cdots, L-1 \rbrace, \ q=\lbrace 1, \cdots, M-M'_P \rbrace.
\end{cases} 
\end{split}
\end{align}
Since the root Zadoff-Chu sequences \eqref{eq:gamma} are orthogonal, we use the entries of $\mathbf{\Gamma}^H(k)$ as pilots, i.e.
\begin{align} \label{eq:pilot symbols}
\mathbf{U}(k)=\frac{1}{\sqrt M} e^{\mathrm{j}\frac{\pi}{M}k^2}, \ k=0, \cdots, M'_P-1. 
\end{align}
The inter-carrier interference (ICI) induced due to the Doppler shifts is not significant ($\vartheta_{\max}^{\text{com}}\ll \Delta f$) and, therefore, $\mathbf{H}_f$ is nearly diagonal. Thus, we estimate the channel in \eqref{eq:OCDM DFT} using pilot symbols by least squares (LS) method as
\begin{align} \label{eq:LS estimate}
[\hat{\mathbf{H}}_f(kL+l)]_n=\frac{\mathbf{Y}_f^{\text{com}}(kL+l,n)}{\mathbf{U}(k)}, \ l=0, \ k=0, \cdots, M'_P-1.
\end{align}
The channel estimates $\hat{\mathbf{H}}_f(m)$ at the data subcarrier $m$ within any adjacent subcarrier group, $kL<m<(k+1)L$, $\forall n$ is then obtained using linear interpolation as 
\begin{align} \label{LS interpolation}
\hat{\mathbf{H}}_f(kL+l)=\left[ \mathbf{H}_f(k+1)-\mathbf{H}_f(k) \right]\frac{l}{L}+\mathbf{H}_f(k), \ 0<l<L.
\end{align}

We remove the pilots from \eqref{eq:OCDM DFT} corresponding to the $k$ indices. Note that the received symbol matrix after pilot removal is $\hat{ \mathbf{Y}}_f^{\text{com}} \in \mathbb C^{(M-M'_P) \times N}$. Using frequency-domain channel estimates for equalization, the zero forcing (ZF) and the minimum mean square error (MMSE) estimates are, respectively,
\begin{align} \label{eq:equalized symbol}
\begin{split}
\mathbf{\hat{Z}}_{\text{ZF}}&=\hat{\mathbf{H}}_f^{-1} {\mathbf{Y}}_f^{\text{com}}, \\
\mathbf{\hat{Z}}_{\text{MMSE}}&= ( \hat{\mathbf{H}}_f \hat{\mathbf{H}}_f^H+\frac{N_0}{P_s}\mathbf{I}_{M-M'_P})^{-1}{\mathbf{Y}}_f^{\text{com}}.
\end{split}
\end{align}
From \eqref{eq:DFT submatrix}, the input symbols $\hat{\mathbf{X}} \in  \mathbb C^{(M-M'_P) \times N}$ across the data sub-carriers are
\begin{align}
\label{eq:estimated symbol}
\begin{split}
\hat{\mathbf{X}}&=\mathbf{Q}_{M-M'_P}\hat{\mathbf{Z}}_{\text{e}}, \ \text{e}\in \lbrace \text{ZF}, \text{MMSE} \rbrace, 
\end{split}
\end{align}
where $\mathbf{Q}_{M-M'_P}=\mathbf{\Gamma}(kL+l)\mathbf{F}_M^{-1}(kL+l,1:M-M'_P),$  $k=\lbrace 0, \cdots, M'_P-1 \rbrace, \ l=\lbrace 1, \cdots, L-1 \rbrace.$
\begin{figure*}
 \begin{minipage}{1\textwidth}
  \includegraphics[scale=0.25]{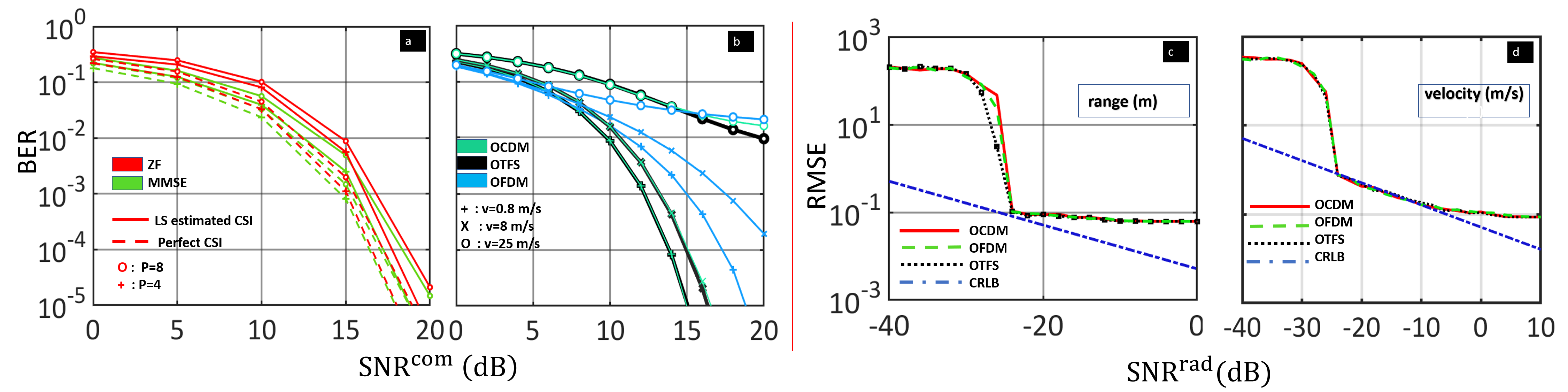}
\caption{OCDM BER performance comparison vs $\text{SNR}^{\text{com}}$ : (a) with LS estimate and perfect CSI, (b) with different waveforms; Root MSE (RMSE) of target parameter estimation  vs $\text{SNR}^{\text{rad}}$: (c) range $\hat{r}$, (d) velocity $\hat{v}$.\vspace{-8pt}
}
\label{fig:Result}
  \end{minipage}
\end{figure*}
\subsection{Target Parameter Estimation}
The received radar signal \eqref{eq:ocdm rad output} after down conversion, sampling and CP removal is
\begin{align} \label{eq:radar samples}
\begin{split}
&\bar{y}^{\text{rad}}_{m,n}= \sum_{p=1}^{P^{\text{rad}}} h_p^{\text{rad}}   e^{\mathrm j2\pi\vartheta_p^{\text{rad}} (nT_0+m\frac{T}{M})}  \sum\limits_{m'=0}^{M-1} x_{n,m'} e^{\mathrm j\frac{\pi}{4}} \\ & \hspace{1.8cm} e^{-\mathrm j\pi\frac{M}{T^2}\left\lbrace(m-m')\frac{T}{M}-\tau_p^{\text{rad}}\right\rbrace^2}+\bar{w}_{m,n}.
\end{split}
\end{align}
Applying matched filtering to recover the chirps symbols gives 
\begin{align} \label{eq: radar matched filtering}
\begin{split}
&y_{m,n}^{\text{rad}} = \frac{1}{M} \sum \limits_{i=0}^{M-1} \bar{y}_{n,i}^{\text{rad}}\exp (-\mathrm j\frac{\pi}{4}) \exp\left[\mathrm j \frac{\pi}{M}(m-i)^2\right] \\
& \approx \sum \limits_{p=1}^{P^{\text{rad}}} x_{m,n} h_p^{\text{rad}}e^{-\mathrm j\pi M \left(\frac{\tau_p^{\text{rad}}}{T^2}\right)^2}  e^{\mathrm j 2\pi \left (n\vartheta_p^{\text{rad}}T_0 - m\tau_p^{\text{rad}}\Delta f\right)}  +  \tilde{w}_{m,n},
\end{split}
\end{align}
where the approximation follows by letting $\vartheta^{\text{rad}}_{\max} \ll \Delta f $ and $\tau^{\text{rad}}_{\max} \ll T$. Assuming the number of targets has been determined (e.g., via hypothesis testing), we find maximum likelihood (ML) estimates of target parameters $\boldsymbol{\theta}=(\tau^{\text{rad}}, \vartheta^{\text{rad}})$, $\forall p$. Note that the Doppler shift and the delay are decoupled in \eqref{eq: radar matched filtering}, which makes the joint estimation problem of $\boldsymbol{\theta}$ simpler. Now, the data symbols in \eqref{eq: radar matched filtering} serve no purpose in estimation of $\boldsymbol{\theta}$ and are thus removed by element-wise division with the estimated symbols $\hat{x}_{m,n}$ \eqref{eq:estimated symbol}. Let $\tilde{y}^{\text{rad}}_{p,m,n}=\frac{y^{\text{rad}}_{p,m,n}}{\hat{x}_{m,n}}$, $\forall p$, where $y^{\text{rad}}_{m,n}=\sum \limits_p y^{\text{rad}}_{p,m,n}$ \eqref{eq: radar matched filtering}. At high SNR, $\hat x_{m,n} \approx x_{m,n}$, implying $\hat x_{m,n}^{\ast} x_{m,n}=|x_{m,n}|^2$. Note that the noise term  $\frac{\tilde{w}_{m,n}}{\hat x_{m,n}}$ \eqref{eq: radar matched filtering} is still AWGN. Thus, the simplified log-likelihood function $\forall p$ is
\begin{align} \label{eq:radar estimate}
\begin{split}
 \scalebox{0.95}{$l(\tilde{\mathbf{Y}}_p^{\text{rad}}|\boldsymbol {\theta},\hat{\mathbf {X}})= 2 h_p^{\text{rad}} \mathbb{Re}\left[\sum \limits_{m,n} 
 \tilde{y}^{\text{rad}}_{p,m,n} e^{-j2\pi n\vartheta_p^{\text{rad}}T_0}e^{ j2\pi m \tau_p^{\text{rad}}\Delta f} \right]-(h_p^{\text{rad}})^2$}.
 \end{split}
\end{align}
Clearly, \eqref{eq:radar estimate} is a two-dimensional (2D) complex periodgram, evaluated as \cite{braun2014ofdm}
\begin{align} \label{eq:periodgram}
\mathbf{\mathcal{Z}}_p(m', n')=\sum \limits_{m=0}^{M_{\text{Per}}-1} \sum \limits_{n=0}^{N_{\text{Per}}-1} \tilde{y}^{\text{rad}}_{p,m,n} e^{-j2\pi \frac{nn'}{N_{\text{Per}}}}e^{ j2\pi \frac{m m'}{   M_{\text{Per}}}},
 \end{align}
where $m'=0, \cdots, M_{\text{Per}}-1$ and $n'=-\frac{N'}{2}, \cdots, \frac{N_{\text{Per}}}{2}-1$ with $M_{\text{Per}}>M$ and $N_{\text{Per}}>N$. 
The target parameters $\boldsymbol \theta$=($\tau^{\text{rad}}$, $\vartheta^{\text{rad}}$), are now estimated following the steps summarized in Algorithm~\ref{algo1}. 
\begin{algorithm}[H]
	\caption{\textit{S}eq\textit{u}e\textit{n}tial symbol \textit{d}ecoding and p\textit{a}rameter \textit{e}stimation (SUNDAE)}
	\label{algo1}
	\begin{algorithmic}[1]
		\Statex \textbf{Input:} Observations $ \bar{y}^{\text{com}}$, $ \bar{y}^{\text{rad}}$
		\Statex \textbf{Output:} Estimated values of $ \left(\hat{\mathbf{X}},\hat{\mathbf{r}}^{\text{rad}}, \hat{\mathbf{v}}^{\text{rad}} \right)$
		\State Estimate the channel frequency response $\mathbf{H}_f$ using the pilot symbols $\hat{\mathbf{ X}}$ in \eqref{eq:LS estimate}
		\State Decode the symbols $\hat{x}_{m,n}$ $\forall n, m$ using \eqref{eq:estimated symbol}
		\State {Use the estimated symbols $\hat{x}_{m,n}$ in \eqref{eq:estimated symbol} for matched filtering of received radar signal $\bar{y}^{\text{rad}}_{m,n}$ in \eqref{eq: radar matched filtering}}
		\For{$p=1:P^{\text{rad}}$}
		\State{Find $(\hat{m},\hat{n})=\arg \max \limits_{(m',n')} |\mathbf{\mathcal Z}_p|^2$ using \eqref{eq:periodgram}; compute $\hat{\vartheta}_p^{\text{rad}}=\frac{\hat{n}}{T_0 N_{\text{Per}}}$ and $\hat {\tau}_p^{\text{rad}}=\frac{\hat{m}}{\Delta f M_{\text{Per}}}$;} 
		\EndFor
		\State\Return{$(\mathbf{\hat X}, \ \hat{\mathbf{r}}^{\text{rad}}=c\hat{\boldsymbol{\tau}}^{\text{rad}}, \ \hat{\mathbf{v}}^{\text{rad}}=c\frac{\hat{\boldsymbol{\vartheta}}^{\text{rad}}}{f_c})$}
	\end{algorithmic}
\end{algorithm}

The Cram\'{e}r-Rao lower bound (CRLB) on the mean squared error (MSE) of the estimates $\boldsymbol{\theta}$ at high SNR is \cite{gaudio2020effectiveness}
\begin{align} \label{eq:CRLB}
    \begin{split}
        \sigma_{\vartheta^{\text{rad}}}^2\geq \frac{6}{(2\pi)^2MN(N^2-1)\text{SNR}^{\text{rad}}}\\
        \sigma_{\tau^{\text{rad}}}^2\geq \frac{6}{(2\pi)^2MN(M^2-1)\text{SNR}^{\text{rad}}}.
    \end{split}
\end{align}
\setlength{\textfloatsep}{0pt}

\vspace{-12pt}
\section{Numerical Experiments}
\label{sec:numexp}
We validated our proposed ISAC scheme based on OCDM through numerical simulations. We set the system parameters to the following values: $f_c=79$ GHz, $B=100$ MHz, $M=256$, $N=50$, $\Delta f=390.63$ KHz, $T_{\text{cp}}=\frac{1}{4}T$, $P^{\text{com}}=3$, $P^{\text{rad}}=1$, $M'_P=4$. The true target range and Doppler velocity are $r^{\text{rad}}= 20$ m and $v^{\text{rad}} = 22.22$ m/s. The input symbols are QPSK-modulated.

Fig.~\ref{fig:Result} (a) compares the BER performance of OCDM for two channel estimation procedures (our LS method and perfect channel state information (CSI)) and two equalizers (ZF and MMSE). We observe that the loss in performance because of linear-interpolation-based channel estimation is within $1$-$2$ dB in both cases. Fig. \ref{fig:Result} (b) compares the BER of our OCDM based scheme against the CP-OFDM and CP-OTFS~\cite{hashimoto2021channel} waveforms, considering perfect CSI. Interestingly, OCDM and OTFS offer nearly identical performance, while OFDM performs the worst. Also, the BER deteriorates as the relative velocity between the vehicles increases and induces greater ICI. 
Figs.~\ref{fig:Result} (c) and (d) depict the root MSE (RMSE) of range and velocity for target parameter estimation as a function of the radar SNR, when the communications signal SNR is set to $15$ dB. For all waveforms, we observe that our SUNDAE algorithm yields nearly identical RMSE, which also matches well with the CRLB in~\eqref{eq:CRLB} at high SNR regimes. 
\section{Summary}
\label{sec:summ}
\vspace{-8pt}
We investigated OCDM as a potential waveform for integrated sensing and communications in automotive scenarios. The proposed design is readily integrated with legacy OFDM systems with a simple FFT-based pre- and post-processing at Tx and Rx. Numerical results highlight that OCDM provides superior communications performance in time-frequency selective channels compared to OFDM under low ICI conditions, while its performance is at par with OTFS with much lower computational overhead. Furthermore, satisfactory bi-static radar sensing performance is attained by our SUNDAE algorithm. This indicates that OCDM is a viable alternative to the existing ISAC multicarrier waveforms. Our practical transceiver structure developed in this paper is useful in exploring several exciting directions for future research.

\clearpage
\bibliographystyle{IEEEtran}
\bibliography{ref_conf}	
	
\end{document}